\documentclass{article}
\usepackage{spconf, amsmath,graphicx}
\usepackage{amssymb, CJK, indentfirst, balance, appendix,booktabs}
\usepackage{threeparttable,subfig}

\title{BA-MoE: Boundary-Aware Mixture-of-Experts Adapter for Code-Switching Speech Recognition}
\name{
\begin{tabular}{c}
\it Peikun Chen$^{1}$, Fan Yu$^{1}$, Yuhao Liang$^{1}$, Hongfei Xue$^{1}$, Xucheng Wan$^{2}$, Naijun Zheng$^{2}$, \\
\it Huan Zhou$^{2}$,  Lei Xie$^{1*}$\thanks{*Corresponding author.}
\end{tabular}
}
\address{$^1$Audio, Speech and Language Processing Group (ASLP@NPU), \\ Northwestern Polytechnical University, Xian, China \\
  $^2$IT Innovation and Research Center, Huawei Technologies}
\copyrightnotice{979-8-3503-0689-7/23/\$31.00~\copyright2023 IEEE}
\begin{document}
%
\maketitle
\begin{abstract}
Mixture-of-experts based models, which use language experts to extract language-specific representations effectively, have been well applied in code-switching automatic speech recognition. However, there is still substantial space to improve as similar pronunciation across languages may result in ineffective multi-language modeling and inaccurate language boundary estimation. To eliminate these drawbacks, we propose a cross-layer language adapter and a boundary-aware training method, namely Boundary-Aware Mixture-of-Experts (BA-MoE). Specifically, we introduce language-specific adapters to separate language-specific representations and a unified gating layer to fuse representations within each encoder layer. Second, we compute language adaptation loss of the mean output of each language-specific adapter to improve the adapter module's language-specific representation learning. Besides, we utilize a boundary-aware predictor to learn boundary representations for dealing with language boundary confusion. Our approach achieves significant performance improvement, reducing the mixture error rate by 16.55\% compared to the baseline on the ASRU 2019 Mandarin-English code-switching challenge dataset.

\end{abstract}
\section{Introduction}
\label{sec:intro}

In recent years, there have been remarkable advancements in deep learning, leading to the widespread adoption of neural end-to-end (E2E) frameworks~\cite{prabhavalkar2023end}, including AEDs~\cite{chorowski2015attention, vaswani2017attention} and neural transducers (NT)~\cite{graves2012sequence}, within the field of ASR. 
Code-switching, the act of alternating between two or more languages in a single sentence, has become increasingly common in today's globalized and culturally diverse world~\cite{moyer2002bilingual}. This linguistic phenomenon presents a significant challenge for speech and language processing tasks, particularly ASR. 
Code-switching automatic speech recognition (ASR) has been extensively studied, initially within the traditional hybrid ASR paradigm~\cite{guo2018study}. Subsequently, various 
 E2E approaches have been proposed, leading to notable progress~\cite{li2019towards,yan2023towards,dalmia2021transformer,song2022language,shi2020asru,shah2020first}.

The major challenges of code-switching ASR consist of the following two aspects: the efficient modeling of language-specific representation  and the accurate prediction of language boundaries. Specifically, effective modeling of multiple languages simultaneously in a unified neural architecture is a major challenge due to the differences in the modeling units of different languages, despite their similarities in pronunciation~\cite {yan2023towards,sitaram2019survey}. Additionally, the confusion of language boundaries during code-switching can misdirect the model's language recognition tendency, consequently reduce the model performance~\cite{zhang2022reducing,fan2023language}.

Language expert modules have been commonly employed to tackle the first challenge of capturing language-specific knowledge~\cite{yan2023towards, zhang2019towards, zhou2020multi, lu2020bi, tian2022lae, yan2022joint}. However, the bi-encoder approach~\cite{lu2020bi} decomposed network parameters into separate language-specific encoders, which results in a lack of interaction between the encoders and overlooks the linguistic common representation. To overcome this limitation, the language-aware encoder (LAE)~\cite{tian2022lae} introduced a multilingual encoder layer before the top-level monolingual encoder, enabling efficient modeling of representations common across languages. Additionally, Yan et al.~\cite{yan2022joint} proposed a conditionally factorized joint framework for integrating monolingual and code-switch sub-tasks. 
However, these approaches compromise the common representation between the two languages within the knowledge space and limit their interactivity, as they partition code-switched speech into distinct components.

Several approaches have been proposed to implicitly learn language boundary representation to tackle the challenge of boundary confusion~\cite{zhang2022reducing,fan2023language, peng2022internal,song2022monolingual,dong2020cif,yu2021boundary}. For instance, Zhang et al.~\cite{zhang2022reducing} employed a language-based correlated attention mechanism, which is computed independently for each monolingual language within the self-attentive layer of the decoder.
Moreover, Fan et al.~\cite{fan2023language} leveraged the continuous integrate-and-fire (CIF)~\cite{dong2020cif,yu2021boundary} mechanism to predict boundaries by utilizing a monolingual weight estimator. 
Additionally, the Internal Language Model Estimation (ILME)~\cite{peng2022internal} approach integrated language models directly into the model architecture, eliminating the need for shallow fusion techniques. 
However, these approaches only implicitly learn boundary representations, leading to ambiguous boundary representations.

In this paper, we propose the \textbf{B}oundary-\textbf{A}ware \textbf{M}ixture-\textbf{o}f-\textbf{E}xperts (BA-MoE) model, which combines an MoE-Adapter, a cross-layer language adaptation training method, and a boundary-aware training method to address these challenges.
To address the first challenge, we depart from previous approaches that only combine language-specific representations at the end of the encoder~\cite{lu2020bi,tian2022lae,yan2022joint}. Instead, our MoE-Adapter approach incorporates adapters~\cite{hou2021exploiting} to extract language-specific representations at each encoder layer. These representations are then combined through the gating network, allowing for more fine-grained learning of both common and specific representations.
Furthermore, we propose the cross-layer language adaptation training method, aiming to improve the learning of language-specific representations for the lower-level adapter modules.
To address the second challenge of language boundary confusion, we propose a boundary-aware training method to combine both implicit and explicit boundary representations. Specifically, we first introduce boundary tokens that explicitly indicate language boundaries. Additionally, we incorporate a boundary-aware predictor that leverages multi-headed self-attention pooling~\cite{lin2017structured, zhu2018self} on the encoder output. This allows the model to focus on relevant, boundaries split segment-level representations for each language. Furthermore, we utilize segment-level acoustic representations and explicit speech boundary multiplexing ASR decoder for joint training.

Experimental results conducted on the ASRU 2019 Mandarin-English code-switching challenge dataset~\cite{shi2020asru} demonstrate the superiority of our proposed model over the baseline model, achieving a relative decrease of 16.55\% and 40.81\% in the mixture error rate and boundary-switching error rate compared to the baseline. 

\section{Method}
\label{sec:Method}
\begin{figure}[h]
\centering
\setlength\abovedisplayskip{0cm}
\setlength\belowdisplayskip{0cm}
\includegraphics[width=\linewidth]{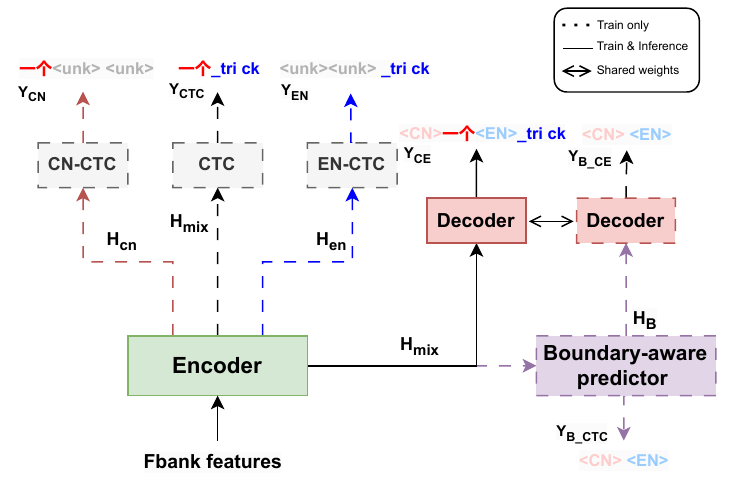}
\caption{Architecture of BA-MoE (English translation of input utterance:``One trick")}
\label{fig: BAMOE}
\end{figure}

  In our approach, we utilize the MoE-Adapter as the encoder to extract speech representations. To facilitate monolingual adaptation, we employ a cross-layer language adaptation training method to optimize the adapter module. The attentive pooling mechanism is employed as the boundary-aware predictor, enabling the prediction of segment-level language switching. Finally, we employ a Transformer decoder to predict both the generated labels and boundary tokens for frame-level and segment-level acoustic representations, respectively. The model details are presented in Fig.~\ref{fig: BAMOE}.

\begin{figure*}[thp]
\centering
\resizebox{0.90\textwidth}{!}{\includegraphics{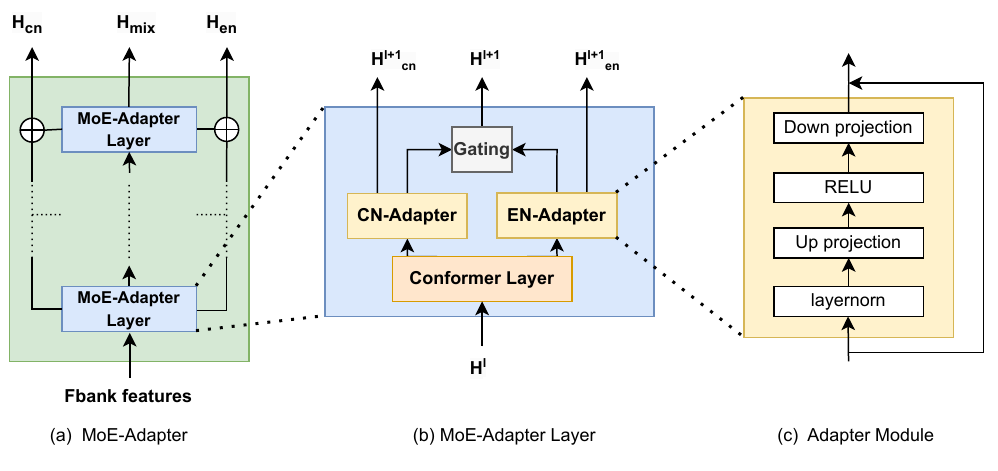}}
\vspace{-0.2cm}
\caption{(a) MoE-Adapter, (b) MoE-Adapter Layer, (c) Adapter module.}
\label{fig: MOE}
\vspace{-0.2cm}
\end{figure*}

\subsection{MoE-Adapter}
\label{ssec:subhead}
To effectively capture language-specific knowledge between different languages, our approach avoids entirely separate modeling for each language. We have incorporated the adapter module as the expert module to capture language-specific representations. As depicted in Fig. 2, the Adapter module consists of layer normalization, an upper projection layer, a non-linear activation function, a lower projection layer, and a residual connection~\cite{HeZRS16} that preserves the original representation. In contrast to other language expert modules, our approach introduces a gating network after the adapters of each layer. This integration facilitates better assimilation of language representations. Consequently, each MoE-Adapter Block contains a Conformer layer~\cite{gulati2020conformer}, followed by separate access to the Chinese and English adapters and a gating network. The outputs of adapters are combined using a gating network, generating the output of the MoE-Adapter Block. Subsequently, this output is passed to the next block for further processing.
As depicted in Fig. \ref{fig: MOE}, the functionality of the MoE-Adapter Layer in layer $l$ can be expressed as follows: Initially, the input representations undergo an extraction of common features through the shared Conformer layer. Following this, separate adapter blocks are used to obtain the respective language-specific representations. 
\begin{equation}
\operatorname{Adapter}(\mathbf{A}^l)=\mathbf{A}^l+\mathbf{W}^l_d(\operatorname{ReLU}\left(\mathbf{W}^l_u\left(\operatorname{LN}\left(\mathbf{A}^l\right)\right)\right)),
\label{eq:adapter}
\centering
\end{equation}
where $\mathbf{A}^l$ represents the inputs to the adapter and outputs of conformer in layer $l$; $\mathbf{W}^l_u$ and $\mathbf{W}^l_d$ are the up-sampling and down-sampling layers of the layer $l$.In addition, we incorporate residual connections within the adapters.

These representations are then combined using a gating network to merge the inputs from the two adapter hidden representations, resulting in the desired acoustic representations. 
\begin{equation}
\mathbf{H}^{l+1}=\operatorname{Gate}(\operatorname{Adapter}_{cn}(\mathbf{A}^l),\operatorname{Adapter}_{en}(\mathbf{A}^l)),
\label{eq:adapter_OUT}
\centering
\end{equation}
where $\mathbf{H}^l$ represents the outputs of layer $l$; $\operatorname{Gate}$ is a linear layer used to learn the weight coefficients at the frame level for both languages.
Subsequently, the processed representations are passed to the next block. We utilize the final encoder layer of $\mathbf{H}^l$ as $\mathbf{H}_{mix}$.

\subsection{Cross-layer language adaptation training}
\label{sssec:subsubhead}

To further improve the adapter's learning capacity for capturing language-specific representations and the correlation between different adapters, we propose a cross-layer language adaptation (CLA) training method. In addition, we utilize the output of multi-layer monolingual Adapters as the monolingual language representation, thereby improving the differentiation between the underlying adapters for the two languages.

\begin{equation}
\vspace{-0.2cm}
\operatorname{H}_{cn}=\frac{1}{L}\sum_{i=1}^{L}\operatorname{Adapter_{cn}}(\mathbf{H}^i),
\label{eq: adapter_loss}
\centering
\end{equation}

\begin{equation}
\operatorname{H}_{en}=\frac{1}{L}\sum_{i=1}^{L}\operatorname{Adapter_{en}}(\mathbf{H}^i),
\label{eq:adapter_LOSS}
\centering
\end{equation}
where $L$ represents the number of adapter layers involved in the calculation. During training, the Mandarin target sequence $Y_{CN}$ is generated by masking the English tokens in the target sequence using the special token \textless Unk\textgreater, while the English target sequence $Y_{EN}$ is generated using the same special token \textless Unk\textgreater. In this section, we employ the CTC loss as an auxiliary criterion, with $Y_{CN}$ and $Y_{EN}$ representing the targets for CN-CTC and EN-CTC. 
\begin{equation}
\mathcal{L}_{\mathrm{CLA}}=\frac{\mathcal{L}_{\mathrm{CN_{CTC}}}+\mathcal{L}_{\mathrm{EN_{CTC}}}}{2},
\centering
\end{equation}
where $\mathcal{L}_{\mathrm{CLA}}$ is the combination of two Losses $\mathcal{L}_{\mathrm{CN_{CTC}}}$ and $\mathcal{L}_{\mathrm{EN_{CTC}}}$, represent the CTC loss for Mandarin Adapter and English Adapter, respectively.

\subsection{Boundary-aware trainng}

To address the boundary confusion challenge, we propose a boundary-aware training (BAT) method. Firstly, we obtain $Y_{att}$ by labeling the targets by utilizing explicit boundary information, whereby \textless CN\textgreater and \textless EN\textgreater serve as the token denoting the language boundary. However, directly incorporating these tokens fails to provide sufficient boundary representations to the encoder. To overcome this limitation, we employ a boundary-aware predictor that implicitly predicts language switching based on acoustic representations. Since language-switching representation is not available in all frames, we utilize self-attention pooling to map the encoder output from frame-level acoustic representations to segment-level ones. By leveraging the self-attention mechanism, we are able to learn the weights that optimize language switching prediction. The attention mechanism takes the whole $\mathbf{H}_{mix}$ as input, and outputs a vector of weights $\mathbf{A}$:

\begin{equation}
\mathbf{A}=softmax(\operatorname{ReLU}(\mathbf{H}_{mix}\mathbf{W}_1)\mathbf{W}_2),
\centering
\end{equation}
where $\mathbf{W}_1$ is a matrix of size $d \times d_a$; $\mathbf{W}_2$ is a matrix of size $d_a \times d_r$, $d_r$ and $d_a$ is a hyperparameter that represents the number of attention heads and dims. 
\begin{equation}
\mathbf{H}_B=\mathbf{A}^{T}\mathbf{H}_{mix},
\centering
\end{equation}
where $\mathbf{H}_B$ is a segment-level boundary representation. Next, we utilize the representation $\mathbf{H}_B$ to classify each segment: 
\begin{equation}
\mathcal{L}_{\mathrm{B_{CTC}}}=\sum\limits_{\mathcal{Y}^{B}}p(Y^{B}|\mathbf{H}_B),
\centering
\end{equation}
where functions are defined as summations over all possible frame-to-label sequences $Y^{B} \in \mathcal{Y}^{B} (\mathbf{A}^{T}) $respectively.

Furthermore, to improve switch detection at language boundaries, We pass the resulting representations $\mathbf{H}_B$ to a decoder, that shares parameters with the ASR decoder.

Next, we compute the cross-entropy loss function $\mathcal{L}_{B_{CE}}$ for the output of the decoder. The loss function for the Boundary-Aware training is defined as follows:
\begin{equation}
\mathcal{L_B}=\mathcal{L}_{B_{CE}}+\mathcal{L}_{B_{CTC}}.
\centering
\end{equation}

\subsection{Loss function}
\label{sssec:Lossfunction}
During training, all the network parameters are optimized by four loss functions, which are cross-entropy (CE), CTC, CLA loss, and boundary-aware loss. Thus, the final loss is:
\begin{equation}
\vspace{-0.2cm} 
\mathcal{L}=\lambda_{ce}\mathcal{L}_{ce}+\lambda_{ctc}\mathcal{L}_{ctc}+\lambda_{C}\mathcal{L}_{CLA}+\lambda_{B}\mathcal{L}_{B},
\centering
\vspace{-0.1cm} 
\end{equation}
where $\lambda_{ce}$, $\lambda_{ctc}$, $\lambda_{C}$, and $\lambda_{B}$ are tunable hyper-parameters. $\mathcal{L}_{ctc}$ and $\mathcal{L}_{ce}$ are computed based on the predictions generated by the ${Y}_{CTC}$ and ${Y}_{CE}$.

\section{Experimental Setups}
\label{sec:Experiments}

\subsection{Datasets}
We perform experiments with ASRU 2019 Mandarin-English code-switching challenge dataset~\cite{shi2020asru}. The corpus consists of about 200 hours of code-switching training data and 500 hours of monolingual Mandarin training data.  The development set and the test set each consist of 20-hour code-switching data. To facilitate experimental design, we further include the 460-hour subset of Librispeech English dataset~\cite{panayotov2015librispeech} into the training set, same as the setup in~\cite{zhang2022reducing,lu2020bi}. 

\subsection{Model configuration}
 The acoustic feature of 80 dimensions log mel-filter bank (Fbank) is extracted from every frame with a frame length of 25ms and frame shift of 10ms.  For Mandarin, 5173 characters are used as the modeling unit. For English, we use byte pair encoding to generate 5000 subwords as the modeling unit. 

To ensure a fair comparison across different approaches, we maintain equal numbers of parameters when comparing model sizes. This approach allows us to evaluate the performance of different models on an equal footing, eliminating potential biases arising from variations in parameter count. All encoders are stacked Conformer[27] layers, in which the attention dimension, feed-forward dimension, number of attention heads, and number of convolutional kernels are fixed to 256, 1024, 4, and 31 respectively.
Five encoder architectures with similar parameter budgets are designed for comparison:
\textbf{Baseline}: 16 stacked Conformer layers.
\textbf{Gating Conformer~\cite{lu2020bi}}: two separated encoders with 8 Conformer layers stacked each.
\textbf{LAE Conformer~\cite{tian2022lae}}: the shared block contains 8 conformer layers while the language-specific blocks consist of 8 Conformer layers each.
\textbf{Attention Module~\cite{zhang2022reducing}}: 16 stacked Conformer layers with independent self-attention decoders.
\textbf{BA-MoE}: 12 stacked MoE-Adapter layers.
All decoders are 6 stacked Transformer[27] layers, in which the attention dimension, feed-forward dimension, and number of attention heads are fixed to 256, 1024, and 4 respectively.

 We set the $\lambda_{att} = 0.7$, $\lambda_{ctc} = 0.3$, $\lambda_{B} = 0.1$ and $\lambda_{C} = 0.1$ in the training stage. For the BAT module, we set the values of $d_a$ and $d_r$ to 128 and 8, respectively, as the dataset contains a maximum of six language switches.

\subsection{Evaluation metrics}
All our experiments are conducted on WeNet toolkits and char error rate (CER) for Chinese part error rate, word error rate (WER) for English part error rate, and mixture error rate (MER) for mixture part error rate. We define the boundary error rate (BER) as an evaluation index of whether the model can correctly distinguish the language boundaries.
\begin{equation}
\centering
\text{BER}=\frac{Insertions+Substitutions+Deletions}{Total\ \ Correct\ \ boundary\ \ Tokens}
\end{equation}
\section{Results}
\label{sec:Results}
\subsection{Comparison of different approaches}

As shown in Table~\ref{table: result}, we evaluate our approach on ASRU 2019 Mandarin-English code-switching challenge Test set. 
Our approach outperforms the  baseline, leading to 16.55\% (12.32\% $\to$ 10.28\%) relative MER reduction on the Test set, respectively. Our method achieves impressive results with the lowest CER, WER, and MER of 10.28\%, 8.16\%, and 27.48\% respectively in the Test sets compared to other methods. The obtained results clearly demonstrate that our method surpasses other schemes with comparable parameter budgets. This performance superiority establishes our approach as an effective solution for the given task. In terms of boundary  prediction, our proposed model brings 40.81\% (3.97\%$\to$2.35\%) relative BER reductions on the Test set.
\begin{table}[h]
\vspace{-0.2cm} 
\caption{Results for various approaches on the Test set (\%).}
\vspace{-0.2cm} 
\centering
\begin{threeparttable}
\resizebox{0.48\textwidth}{!}{
\begin{tabular}{lccccc}
\toprule
Model           & Params(M)& MER(\%)&    CER(\%)    &      WER(\%) &  BER(\%)     \\ \hline
Baseline        &41 &    12.32   &   10.05       &      30.71   &  3.97      \\
Gating Conformer\tnote{*}~\cite{lu2020bi} &41&    11.30   &   9.01        &      30.01   &  3.67      \\
LAE Conformer\tnote{*}~\cite{tian2022lae}  &41  &    11.14   &   8.88      &      29.55   &  3.58      \\
Attention Module\tnote{*}~\cite{zhang2022reducing} &43 &    10.87   &   8.74 &  28.21   &  3.25    \\ \hline
BA-MoE          &43 &    \textbf{10.28}   &   \textbf{8.16}        &      \textbf{27.48}   &  \textbf{2.35}      \\ 
\bottomrule
\end{tabular}
}
\begin{tablenotes}
    \footnotesize
		\item *: These models are re-implemented by ourselves.
\end{tablenotes}
\end{threeparttable}
\label{table: result}
\vspace{-0.8cm} 
\end{table}

\subsection{Visualization of cross-layer adaption loss}

\begin{figure}[h]
\vspace{-0.4cm}
  \centering 
   \subfloat[without CLA loss]
  {
      \label{fig:subfig1}\includegraphics[width=0.235\textwidth]{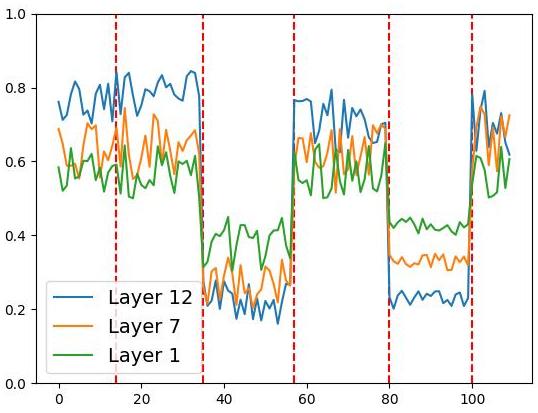}
  }
  \subfloat[with CLA loss]
  {
      \label{fig:subfig2}\includegraphics[width=0.235\textwidth]{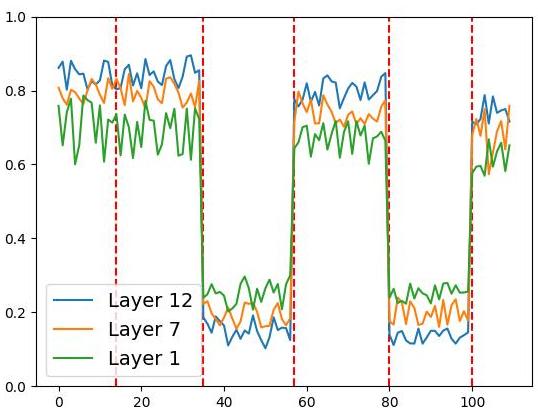}
  }
  \caption{Visualization of  Mandarin gating coefficients learned from CLA loss for the utterance index ASRU-CS-TEST-16146 (the red dotted line represents the language boundary, including the mute paragraph of the head and tail).}
  \label{fig:subfig_1}
\vspace{-0.8cm}
\end{figure}

Fig.~\ref{fig:subfig1} and Fig.~\ref{fig:subfig2} illustrate the weight coefficients of the gating networks for Mandarin in Layer 1, Layer 7, and Layer 12, both with and without CLA Loss. As depicted in Fig.~\ref{fig:subfig1}, in the absence of CLA Loss, the adapters have limited learning capacity for language-specific representations. However, in Fig.~\ref{fig:subfig2}, with the inclusion of CLA Loss, the weight coefficients of the layers become more distinct, indicating an improved ability of the adapters to capture language-specific characteristics. This highlights the effectiveness of the CLA Loss in enhancing the adapter's capacity for learning language-specific features.

\subsection{Visualization of boundary-aware learning}
\begin{figure}[h]
\vspace{-0.3cm} 
\centering
\setlength\abovedisplayskip{0cm}
\setlength\belowdisplayskip{0cm}
\includegraphics[width=\linewidth]{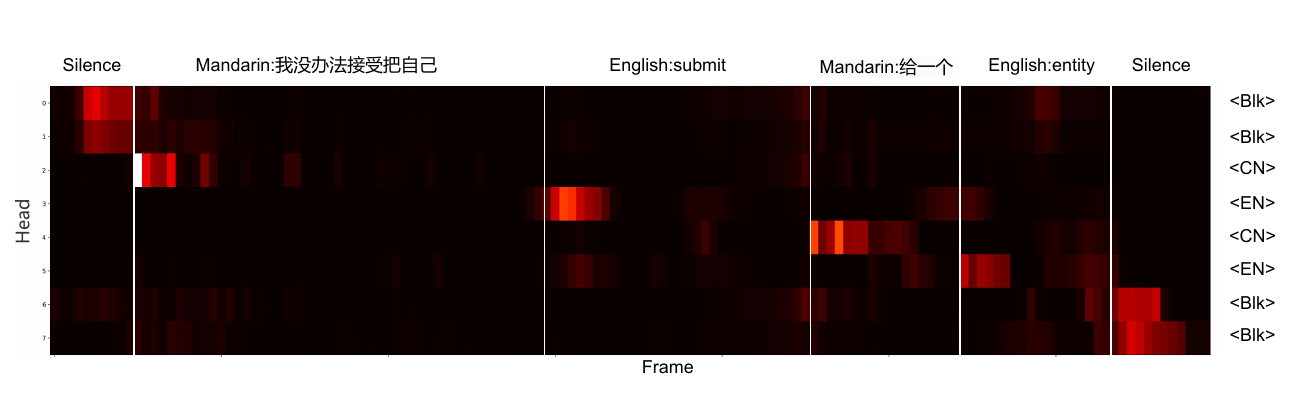}
\vspace{-1cm} 
\caption{Visualization of Boundary-aware learning for the utterance index ASRU-CS-TEST-16122.}
\label{fig: vis}
\vspace{-0.8cm} 
\end{figure}

In Fig.~\ref{fig: vis}, the horizontal axis represents the time range, and the vertical axis represents the attention headings. The white lines indicate the language boundaries enforced by alignment. The graph is generated from the weight matrix $\mathbf{A}^{T}$ computed by the self-attention mechanism. The speech is processed by the boundary-aware predictor, resulting in the representation $\mathbf{H}_B$ composed of speech segments. The classifier generates the sequential result of \textless CN\textgreater\textless EN\textgreater\textless CN\textgreater\textless EN\textgreater same as the ground truth while ignoring \textless Unk\textgreater, which represents a mute segment. Notably, when the speech transitions to a new language, we observe a sudden increase in the weight coefficients, leading to a distinct bright block at the boundary. This observation effectively illustrates the impact of our boundary-aware training.

\subsection{Ablation Study}
We conducted ablation experiments on our proposed method, and the results are summarized in Table~\ref{table: ablation}. Specifically, we performs ablations on boundary-aware learning, adaption loss, and MoE-Adapter structures. The findings reveal that the absence of these components weakens the recognition performance of the model. Notably, the MoE-Adapter has the most significant impact on the overall MER (12.32\%$\to$11.12\%). This underscores the importance of the MoE-Adapter in effectively modeling confounding acoustics in code-switch ASR. Additionally, the CLA loss plays a crucial role in improving the monolingual modeling ability of the adapter, bringing 0.51\% (11.12\%$\to$10.61\%) absolute MER reduction on the Test set. Finally, the results demonstrate that the BAT method successfully achieves 0.33\% (10.61\%$\to$10.28\%) absolute MER reduction by effectively detecting language boundaries.

\begin{table}[h]
\vspace{-0.2cm}
\caption{
    Ablation study on the Test set (\%).
}
\vspace{-0.2cm} 
\centering
\resizebox{0.48\textwidth}{!}{
\begin{tabular}{lcccc}
\toprule
Model                          & MER(\%)&    CER(\%)    &      WER(\%) &  BER(\%) \\ \hline 
Baseline                        &12.32  &   10.05 & 30.71   & 3.97   \\
\ \ + MoE-Adapter           &11.12  &   8.89  & 29.25   & 3.62  \\
\ \ \ \ + CLA loss     &10.61  &   8.41  & 28.56   & 3.56  \\
\ \ \ \ \ \ + BAT           & \textbf{10.28} &   \textbf{8.16}  & \textbf{27.48}  & \textbf{2.35}\\\bottomrule
\end{tabular}
}
\label{table: ablation}
\centering
\vspace{-0.7cm}
\end{table}
\vspace{-0.15cm} 
\subsection{Impact of the model size}
Due to the varying model sizes of previous methods on the ASRU 2019 Mandarin-English code-switching challenge dataset, we increase our model size for a fair comparison with these methods. Our model size fixes the attention dimension, feed-forward dimension, and the number of attention heads to 512, 8, and 1024, respectively. It is worth noting that our model size is similar to the other two methods, as shown in Table~\ref{table:result2}. Meanwhile, we integrate a transformer language model (LM) into our proposed model to improve the language generalization ability, which brings 2.6\% relative MER reductions on Test sets. Additionally, in the final MER, we achieved an absolute reduction of 0.49\%(8.57\%$\to$8.08\%).

\begin{table}[h]
\vspace{-0.2cm} 
\caption{Results of the large model on the Test set (\%).}
\vspace{-0.1cm} 
\centering
\resizebox{0.48\textwidth}{!}{
\begin{tabular}{lcccc}
\toprule
Model      &Params(M)     &      MER(\%)&    CER(\%)    &      WER(\%)     \\ \hline
LAE Conformer~\cite{tian2022lae}  & 138  &    8.9     &   7.3         &      27.7         \\
Attention Module~\cite{zhang2022reducing}&112    &    8.57    &   6.68        &      24.11          \\ \hline
BA-MoE     &125      &    8.30    &   6.46        &      23.26       \\
\ \ + LM   & -       &    \textbf{8.08}    &   \textbf{6.28}        &      \textbf{22.78}        \\\bottomrule
\end{tabular}
}
\label{table:result2}
\vspace{-0.7cm}
\end{table}

\section{Conclusion}
\vspace{-0.2cm} 
In this work, we propose Boundary-Aware Mixture-of-Experts (BA-MoE), an approach that effectively models language-specific representation and incorporates boundary-aware learning. To address the challenges posed by similarities in pronunciation across languages, we use MoE-Adapter as an encoder that enables the separation and fusion of language-specific representations at a finer-grained level within each encoder layer.  Furthermore, we propose a cross-layer language adaptation training method to improve the adapter module's capability for language-specific modeling. To mitigate language boundary confusion, we utilize a boundary-aware predictor to learn boundary representations for dealing with language boundary confusion. Experimental results on the ASRU test set demonstrates a relative reduction of 16.55\% and 40.81\% in MER and BER, compared to the baseline model.


\bibliographystyle{IEEEbib}
\bibliography{strings,refs}

\end{document}